\documentclass[twocolumn, pra,superscriptaddress]{revtex4-1}
\usepackage{graphicx}
\usepackage{epsfig}
\usepackage{color}
\usepackage{amsmath,bm}
\begin{document}
\title{Universal relations for dilute systems with two-body decays in reduced dimensions}
\author{Mingyuan He}
\email{hemingyuan7@gmail.com}
\affiliation{Shenzhen JL Computational Science and Applied Research Institute, Shenzhen 518131, China}
\author{Chenwei Lv}
\affiliation{Department of Physics and Astronomy, Purdue University, West Lafayette, Indiana 47907, USA}
\author{Qi Zhou}
\email{zhou753@purdue.edu}
\affiliation{Department of Physics and Astronomy, Purdue University, West Lafayette, Indiana 47907, USA}
\affiliation{Purdue Quantum Science and Engineering Institute, Purdue University, West Lafayette, Indiana 47907, USA}
\date{\today}
\begin{abstract}

Physical systems in reduced dimensions exhibit intriguing properties. For instance, the dependences of two-body and many-body physics on scattering lengths are distinct from their counterparts in three dimensions. Whereas many studies of ultracold atoms and molecules in reduced dimensions have been focusing on closed systems, two-body losses may occur in such systems.  Here, we show that the two-body inelastic loss rate in reduced dimensions can be expressed in universal relations that are governed by contacts. These universal relations correlate the two-body decay rate with other physical observables at zero and finite temperatures and generic interaction strengths. Our results will provide experimentalists with a new protocol to study inelastic scatterings in both few- and many-body systems in reduced dimensions. 

\end{abstract}

\maketitle

\section*{I. INTRODUCTION}

Ultracold atoms have provided physicists with a highly tunable platform to explore quantum few-body and many-body systems in both three dimensions (3D) and reduced dimensions \cite{Posazhennikova2006,Zwerger2008,Stringari2008,Chin2010}. In reduced dimensions, microscopic parameters control few-body and many-body physics in distinct means. For instance, the dependence of phase shifts on energies in one dimension (1D) and two dimensions (2D) are distinct from that in 3D \cite{Olshanii1998,Petrov2001}. Furthermore, in the celebrated universal relations that underlie ultracold atoms and other dilute quantum systems in arbitrary dimensions, the s-wave scatterings enter these relations in terms of $a_0$ and $\ln a_0$ in 1D and 2D, respectively, while universal relations in 3D often include $1/a_0$ \cite{Tan2008a,Tan2008b,Tan2008c,Platter2008,Leggett2009,Drut2011,Barth2011,Valiente2011,Valiente2012,Hofmann2012a,Langmack2012,Castin2012a,Castin2012b,Vignolo2013,Hofmann2014a,Chen2014a,Weiss2015a,cDrut2015,Patu2017,Decamp2018,He2019,Bougas2021,Sekino2021}. $a_0$ is the s-wave scattering length. It is thus an important task to find the counterparts of 3D universal relations in reduced dimensions, which helps us to study how contacts manifest themselves in reduced dimensions. In such studies of universal relations, most works have been focusing on systems with elastic scatterings. However, two-body losses due to inelastic collisions may occur in realistic systems \cite{Julienne1999}. It is thus desirable to explore how inelastic scatterings may change the universal relations or provide us with conceptually new relations in 3D and reduced dimensions.

In addition to ultracold atoms, ultracold molecules have also been well established as a powerful platform to study a wide range of important topics  in condensed matter physics, atomic, molecular and optical physics, and chemical physics \cite{Ye2017a,Ye2017b,Fitch2021,Ni2021rev}. One of the key issues emerged in experiments is the inelastic loss of molecules \cite{Deiglmayr2008,Ye2010,Nagerl2014,Cornish2014,Zwierlein2015,Wang2016,Rvachov2017,Wang2018,Ye2019,Gregory2019,Ni2019,Ni2020,Doyle2020,Ye2020,Yang2020,Wang2021,Luo2022,Gregory2020,Wang2021b}. For instance, two reactive molecules can get close and react as ${\rm AB}+{\rm AB} \to {\rm A_2} + {\rm B}_2$, which leads to the loss of ${\rm AB}$ molecules \cite{Julienne2010,Ye2019}. Even in the absence of reactions, the formation of complexes could also lead to two-body decays \cite{Wang2018,Gregory2019,Mayle2012,Mayle2013,Christianen2019a,Christianen2019b,Rey2020}. Similar to atoms, ultracold molecules can also been prepared in reduced dimensions \cite{Ye2020b,Ye2020c,Chin2021,Rosenberg2021,Ye2021a,Gohle2018,Luo2021}.  A recent pioneering experiment has made an attempt to explore how the two-body decay may change with reducing the dimension by increasing the transverse confinement \cite{Ye2020b}.

In 3D,  it has been recognized that universal relations exist in systems with two-body losses \cite{Hammer2013,Hammer2017,He2020}. Such relations directly correlate two-body decays with other many-body properties such as the momentum distribution and the density-density correlation function. Motivated by the importance of studying ultracold atoms and ultracold molecules in reduced dimensions, in this manuscript, we explore universal relations in 1D and 2D systems when an arbitrary partial wave scattering is inelastic. We show that the two-body inelastic loss rate can be expressed as contacts multiplied by microscopic parameters determined purely by two-body physics at short range, similar to those obtained in 3D. As such, our results are valid at zero and finite temperatures and generic interaction strengths and will provide experimentalists a useful protocol to explore two-body decays in a many-body environment in reduced dimensions. 

The rest of this paper is organized as follows. In Sec. II, we provide a generic method of deriving the two-body inelastic loss rate, the momentum distribution, and the density correlation function in $d$-dimensional ($d$D) systems, where $d=1,2,3$. In Sec. III, we consider single-component ultracold reactive molecules in 1D and derive the exact relations between contacts and physical quantities including the two-body inelastic loss rate, the momentum distribution, and the density correlation function. Similar discussions for 2D are given in Sec. IV. Furthermore, we discuss the temperature dependence of the loss rate in both the homogeneous systems and the harmonic traps in Sec. V. We conclude our results in Sec. VI.

\section*{II. TWO-BODY INELASTIC LOSS RATE FOR REACTIVE MOLECULES IN $d$ DIMENSIONS}
We consider a single-component system of $N$  reactive molecules. The Hamiltonian is written as
\begin{equation}
H=\sum_{i=1}^N [-\frac{\hbar^2}{2M}\nabla_i^2+V_{\rm ext} ({\bf x}_i)] + \sum_{i<j} U({\bf x}_{ij}),
\end{equation}
where $M$ is the mass of each molecule and ${\bf x}_i=(x_i^{(1)},x_i^{(2)},\cdots,x_i^{(d)})$ is the coordinate of the $i$-th molecule in $d$D space. ${\bf x}_{ij}={\bf x}_i-{\bf x}_j=(x_{ij}^{(1)},x_{ij}^{(2)},\cdots,x_{ij}^{(d)})$. $V_{\rm ext} ({\bf x}_i)$ is the external trap. $U({\bf x}_{ij})=U_{\rm R}({\bf x}_{ij})+iU_{\rm I}({\bf x}_{ij})$ is the complex two-body short-range interaction, which captures the two-body inelastic collisions and is nonzero only when $|{\bf x}_{ij}|<r_0$. $U_{\rm I}({\bf x}_{ij})$ is non-positive and nonzero at an even shorter distance characterized by $r^*$, $|{\bf x}_{ij}|<r^*<r_0$, where the chemical reaction happens. The many-body wavefunction, which is an eigenstate of the system, satisfies the Schr\"odinger equation
\begin{equation}
i\hbar \frac{\partial }{\partial t} \Psi({\bf x}_1,{\bf x}_2,\cdots,{\bf x}_N) = H \Psi({\bf x}_1,{\bf x}_2,\cdots,{\bf x}_N).
\end{equation}
We consider a finite system, the net current of which vanishes at large distance. The two-body inelastic loss rate is written as
\begin{equation}
\frac{\partial N}{\partial t} = \frac{4}{\hbar} \sum_{i<j} \int  \prod_{i=1}^N {\rm d} {\bf x}_i U_{\rm I}({\bf x}_i-{\bf x}_j) \left|\Psi({\bf x}_1,{\bf x}_2,\cdots,{\bf x}_N)\right|^2,\label{decayd}
\end{equation}
which is consistent with the second quantization form using bosonic (fermionic) operators,
\begin{equation}
\frac{\partial N}{\partial t} = \frac{2}{\hbar}  \int   {\rm d} {\bf x}{\rm d} {\bf x}' U_{\rm I}({\bf x}-{\bf x}') \left\langle\Psi^\dag({\bf x})\Psi^\dag({\bf x}')\Psi({\bf x}')\Psi({\bf x})\right\rangle,
\end{equation}
which can be derived from the Lindblad master equation \cite{He2020}.

It is clear that a length scale separation exists in ultracold reactive molecules, i.e., the range of interaction $r_0$ is much shorter than the average inter-particle distance characterized by the inverse of the Fermi momentum $k_F$ while the reactive collisions happen in an even shorter distance characterized by $r^*$, $r^*< r_0\ll k_F^{-1}$. When the distance between any two molecules is much shorter than the average inter-particle distance, i.e., $|{\bf x}_{ij}|\ll k_F^{-1}$, the possibility of a third molecule to get close to these two molecules and interact together at short distance is negligible. It is thus sufficient to consider only the two-body effect. The many-body wavefunction has the asymptotical behavior at short distance, which is stated as
\begin{equation}
\Psi( {\bf x}_1,{\bf x}_2,\cdots,{\bf x}_N)  \stackrel{|{\bf x}_{ij}|\ll k_F^{-1}}{\xrightarrow{\hspace*{1.1cm}} }  \sum_{s,\epsilon} \psi_s({\bf x}_{ij};\epsilon)G_s({\bf X}_{ij};E-\epsilon), \label{asyd}
\end{equation}
where $\epsilon$ is the two-body collision energy and $\psi_s({\bf x}_{ij};\epsilon)$ is the two-body relative wavefunction which satisfies 
\begin{equation}
\label{TBSE}
 [-\frac{\hbar^2}{M}\nabla^2_{{\bf x}_{ij}} + U({\bf x}_{ij})] \psi_s({\bf x}_{ij};\epsilon)=\epsilon \psi_s({\bf x}_{ij};\epsilon).
\end{equation}
$s$ is the angular momentum quantum number, which denotes $(l,m)$ for 3D and $l$ for 1D and 2D. $\nabla^2_{{\bf x}_{ij}}=\sum\nolimits_{n=1}^{d}[\partial^2 /\partial (x_{ij}^{(n)})^2]$. ${\bf X}_{ij}=\{({\bf x}_{i}+{\bf x}_{j})/2,{\bf x}_{k\neq i,j}\}$ denotes the center of mass coordinate of the $i$-th and the $j$-th molecules and the coordinates of all the other $N-2$ molecules. $G_s({\bf X}_{ij};E-\epsilon)$ is the many-body wavefunction, which characterizes the center of mass motion of the $i$-th and the $j$-th molecules and the motions of all the other $N-2$ molecules. 

Whereas $G_s({\bf X}_{ij};E-\epsilon)$ is usually very complex and hard to know, $\psi_s({\bf x}_{ij};\epsilon)$ has a universal asymptotic form when $r_0\leq|{\bf x}_{ij}|\ll k_F^{-1}$. $\psi_s({\bf x}_{ij};\epsilon)=\varphi_s(|{\bf x}_{ij}|;\epsilon) Y_s (\hat {\bf x}_{ij})$, where $\hat {\bf x}_{ij}={\bf x}_{ij}/|{\bf x}_{ij}|$ and $Y_s (\hat {\bf x}_{ij})$ is the generalized spherical harmonics in $d$D. Furthermore, $\varphi_s(|{\bf x}_{ij}|;\epsilon)$ can be expanded as $\varphi_s(|{\bf x}_{ij}|;\epsilon)=\varphi_s^{(0)}(|{\bf x}_{ij}|)+\varphi_s^{(1)}(|{\bf x}_{ij}|) q_\epsilon^2 + {\rm O} (q_\epsilon^4)$, where $q_\epsilon=(M\epsilon/\hbar^2)^{1/2}$. Equation (\ref{asyd}) can then be written as
\begin{equation}
\begin{split}
\Psi( {\bf X}_{ij},{\bf x}_{ij})  \stackrel{|{\bf x}_{ij}|\ll k_F^{-1}}{\xrightarrow{\hspace*{1.1cm}} } & \sum_s \Big[\varphi_s^{(0)}(|{\bf x}_{ij}|)g_s^{(0)}({\bf X}_{ij})\\
&+\varphi_s^{(1)}(|{\bf x}_{ij}|)g_s^{(1)}({\bf X}_{ij})\Big]Y_s (\hat {\bf x}_{ij}),
\end{split}\label{asyd1}
\end{equation}
where $g_s^{(m)}({\bf X}_{ij})=\sum\nolimits_{\epsilon} q_\epsilon^{2m} G_s({\bf X}_{ij};E-\epsilon)$. Starting from Eq. (\ref{asyd1}) and the exact universal asymptotic form of $\varphi_s^{(0)}(|{\bf x}_{ij}|)$ and $\varphi_s^{(1)}(|{\bf x}_{ij}|)$ at $r_0\leq|{\bf x}_{ij}|\ll k_F^{-1}$, a number of universal relations determined by contact $C$, a fundamental quantity in dilute quantum systems, can be derived. $C \sim \int {\rm d} {\bf X}_{ij} g_s^{(\nu)\ast} g_{s'}^{(\nu')}  $. In this manuscript, we will  focus on the universal relations of the two-body inelastic decay rate, the momentum distribution and the density correlation function.

\noindent {\it Two-body inelastic loss rate:} The two-body inelastic loss rate can be obtained by solving Eq. (\ref{decayd}). Starting from Eqs. (\ref{asyd}-\ref{asyd1}), the right-hand side of Eq. (\ref{decayd}) can be obtained by solving the following equation, which is
\begin{equation}
\begin{split}
& {\cal J}\left[U_I({\bf x}_{ij}) \left| \Psi ( {\bf X}_{ij}, {\bf x}_{ij})\right|^2\right] \\
=&\frac{1}{2i}  {\cal J}\Big[\Psi^* ( {\bf X}_{ij}, {\bf x}_{ij}) \sum_{s,\epsilon} \epsilon G_s ({\bf X}_{ij};E-\epsilon) \psi_s ( {\bf x}_{ij};\epsilon)\Big] \\
-& \frac{1}{2i}  {\cal J} \Big[\Psi ( {\bf X}_{ij}, {\bf x}_{ij}) \sum_{s,\epsilon} \epsilon^* G^*_s ({\bf X}_{ij};E-\epsilon) \psi^*_s ( {\bf x}_{ij};\epsilon)\Big] \\
+&\frac{1}{2i}\frac{\hbar^2}{M} {\cal J} \Big[ \Psi^* ( {\bf X}_{ij}, {\bf x}_{ij})\nabla^2_{{\bf x}_{ij}} \Psi ( {\bf X}_{ij}, {\bf x}_{ij}) \\
&\qquad\qquad-\Psi ( {\bf X}_{ij}, {\bf x}_{ij})\nabla^2_{{\bf x}_{ij}} \Psi^* ( {\bf X}_{ij}, {\bf x}_{ij})  \Big],
\end{split}\label{decayTB}
\end{equation}
where ${\cal J}$ is the short-hand notation of the summation and integral $\sum\nolimits_{i<j} \int {\rm d} {\bf X}_{ij} \int_0^{r_0} {\rm d} {\bf x}_{ij}$. The last term on the right-hand side of Eq. (\ref{decayTB}) can be further rewritten as the surface integral,
\begin{equation}
\begin{split}
&\sum_{i<j}\int {\rm d} {\bf X}_{ij} \int_0^{r_0} {\rm d} {\bf x}_{ij}\Big[ \Psi^* ( {\bf X}_{ij}, {\bf x}_{ij})\nabla^2_{{\bf x}_{ij}} \Psi ( {\bf X}_{ij}, {\bf x}_{ij}) \\
&\qquad\qquad\qquad\qquad-\Psi ( {\bf X}_{ij}, {\bf x}_{ij})\nabla^2_{{\bf x}_{ij}} \Psi^* ( {\bf X}_{ij}, {\bf x}_{ij})  \Big]\\
=&\sum_{i<j}\int {\rm d} {\bf X}_{ij} \oint_{|{\bf x}_{ij}|=r_0} \Big[ \Psi^* ( {\bf X}_{ij}, {\bf x}_{ij})\nabla_{{\bf x}_{ij}} \Psi ( {\bf X}_{ij}, {\bf x}_{ij}) \\
&\qquad\qquad\qquad\quad-\Psi ( {\bf X}_{ij}, {\bf x}_{ij})\nabla_{{\bf x}_{ij}} \Psi^* ( {\bf X}_{ij}, {\bf x}_{ij})  \Big] \cdot{\rm d} {\bf S}.
\end{split}\label{decayTB1}
\end{equation}
One can used the mathematics given in Appendix A during the calculation. By taking the explicit expressions of $\varphi_s^{(0)}(|{\bf x}_{ij}|)$ and $\varphi_s^{(1)}(|{\bf x}_{ij}|)$ at $r_0\leq|{\bf x}_{ij}|\ll k_F^{-1}$ into Eq. (\ref{asyd1}) firstly, and bringing Eq. (\ref{asyd1}) back to Eqs. (\ref{decayTB}) and (\ref{decayTB1}) then, Eq. (\ref{decayTB}) can be calculated explicitly. Finally, by taking Eq. (\ref{decayTB}) back to Eq. (\ref{decayd}), the explicit expression of Eq. (\ref{decayd}) can be obtained.

\noindent {\it Momentum distribution:} The momentum distribution can be obtained by using the first quantization form, 
\begin{equation}
n({\bf k}) = \sum_{i=1}^N \int \prod_{j\neq i} {\rm d} {\bf x}_j \left|\int {\rm d} {\bf x}_i \Psi ({\bf x}_1,{\bf x}_2,\cdots, {\bf x}_N)e^{-i{\bf k}\cdot {\bf x}_i}\right|^2.  \label{MDd}
\end{equation}

\noindent {\it Density correlation function:} The density correlation function can be obtained by using the definition
\begin{equation}
\begin{split}
S({\bf x}_{ij})=&\int {\rm d} \frac{{\bf x}_i+{\bf x}_j}{2} \langle n({\bf x}_i)n({\bf x}_j) \rangle \\
=& N(N-1) \int{\rm d} {\bf X}_{ij} \left| \Psi ( {\bf X}_{ij}, {\bf x}_{ij}) \right|^2.
\end{split}\label{DCf}
\end{equation}

Near the Feshbach resonance of a certain partial wave scattering, contacts of this partial wave will be dominant. For simplicity, we consider single-component molecules in a single partial wave scattering channel $s$ in the following discussions. The generalization of these discussions to systems with mixed partial wave scatterings will be straightforward.

\begin{table*}
  \centering
  \caption{The low energy expansion of phase shift $\eta_l$ in different dimensions \cite{Hammer2009}. $\gamma\approx 0.577$ is the Euler's constant. }\label{table1}
  \begin{tabular}{@{}lccccc@{}}
    \hline\hline
     $l$ & One Dimension&\quad\quad& Two Dimensions &\quad\quad& Three Dimensions \\
    \hline
    $l=0$ & $q_\epsilon \tan\eta_0=\frac{1}{a_0}$ && $\frac{\pi}{2}\cot\eta_0= {\rm ln}(\frac{q_\epsilon a_0}{2}e^\gamma)$ && $q_\epsilon\cot\eta_0=-\frac{1}{a_0} + r_0^e q_\epsilon^2$ \\
    $l=1$ & $q_\epsilon \cot\eta_1=-\frac{1}{a_1}+r_1^e q_\epsilon^2$ && $\frac{\pi}{2}q_\epsilon^2 \cot\eta_1-q_\epsilon^2{\rm ln} (\frac{q_\epsilon r_0}{2}e^{\gamma-1/2})=-\frac{1}{a_1} + r_1^e q_\epsilon^2$ && $q_\epsilon^{3}\cot\eta_1=-\frac{1}{a_1} + r_1^e q_\epsilon^2$ \\
    $l>1$ &   && $\frac{\pi}{2} q_\epsilon^{2l}\cot\eta_l=-\frac{1}{a_l} + r_l^e q_\epsilon^2$ && $q_\epsilon^{2l+1}\cot\eta_l=-\frac{1}{a_l} + r_l^e q_\epsilon^2$ \\
    \hline
  \end{tabular}
\end{table*}

\begin{table*}
  \centering
  \caption{The two-body inelastic loss rate in different dimensions. ${\tilde \varphi_{s}^{(0)}}\left( r \right)$ is a wave function obtained from extending the actual wave function ${\varphi_{s}^{(0)}}\left( r \right)$ outside the potential ($r>r_0$) into the regime $r<r_0$. ${\bf R}_{ij}=({\bf R}^{\bm \rho}_{ij},{\bf R}^z_{ij})$.}\label{table2}
  \begin{tabular}{@{}lccc@{}}
    \hline\hline
     & One Dimension ($s=l$) & Two Dimensions ($s=l$) & Three Dimensions \cite{He2020} ($s=lm$)\\
    \hline
    $\partial_t N$ & $ - \frac{2\hbar}{2^2M} \sum\nolimits_{\nu=1}^3\kappa_\nu C_\nu^{(s)}$ & $- \frac{2\hbar}{(2\pi)^2M} \sum\nolimits_{\nu=1}^3\kappa_\nu C_\nu^{(s)}$ & $- \frac{2\hbar}{(4\pi)^2M} \sum\nolimits_{\nu=1}^3\kappa_\nu C_\nu^{(s)}$\\
    $C_1^{(s)}$  & $2^2N(N-1)\int {\rm d}{\bf R}^z_{ij} |g_{s}^{(0)}|^2$  & $(2\pi)^2N(N-1)\int {\rm d}{\bf R}^{\bm \rho}_{ij} |g_{s}^{(0)}|^2$ & $(4\pi)^2N(N-1)\int {\rm d}{\bf R}_{ij} |g_{s}^{(0)}|^2$ \\
    $C_2^{(s)}$ & $2(2^2)N(N-1)\int {\rm d}{\bf R}^z_{ij} {\rm Re}(g_{s}^{{(0)}*}g_{s}^{(1)})$  & $2(2\pi)^2N(N-1)\int {\rm d}{\bf R}^{\bm \rho}_{ij} {\rm Re}(g_{s}^{{(0)}*}g_{s}^{(1)})$ & $2(4\pi)^2N(N-1)\int {\rm d}{\bf R}_{ij} {\rm Re}(g_{s}^{{(0)}*}g_{s}^{(1)})$ \\
    $C_3^{(s)}$  & $2(2^2)N(N-1)\int {\rm d}{\bf R}^z_{ij} {\rm Im}(g_{s}^{{(0)}*}g_{s}^{(1)})$  & $2(2\pi)^2N(N-1)\int {\rm d}{\bf R}^{\bm \rho}_{ij} {\rm Im}(g_{s}^{{(0)}*}g_{s}^{(1)})$ & $2(4\pi)^2N(N-1)\int {\rm d}{\bf R}_{ij} {\rm Im}(g_{s}^{{(0)}*}g_{s}^{(1)})$ \\
    $\kappa_1$ & $-\frac{M}{{{\hbar ^2}}}\int_0^{\infty} {{| \varphi_{s}^{(0)}({r})|^2}U_I({r}){\rm d}r}$ & $-\frac{M}{{{\hbar ^2}}}\int_0^{\infty} {{| \varphi_{s}^{(0)}({r})|^2}U_I({r})r{\rm d}r}$ & $-\frac{M}{{{\hbar ^2}}}\int_0^{\infty} {{| \varphi_{s}^{(0)}({r})|^2}U_I({r})r^2 {\rm d}r}$ \\
    $\kappa_2$ & $-\frac{{M}}{{{\hbar ^2}}}{\mathop{\rm Re}\nolimits} ( \int_0^\infty  { \varphi_{s} ^{(0)*}({r})\varphi_{s} ^{(1)}(r)U_I({r}){\rm d}{r}}  )$ & $-\frac{{M}}{{{\hbar ^2}}}{\mathop{\rm Re}\nolimits} ( \int_0^\infty  { \varphi_{s} ^{(0)*}({r})\varphi_{s} ^{(1)}(r)U_I({r})r{\rm d}r}  )$ & $-\frac{{M}}{{{\hbar ^2}}}{\mathop{\rm Re}\nolimits} ( \int_0^\infty  { \varphi_{s} ^{(0)*}({r})\varphi_{s} ^{(1)}(r)U_I({r})r^2{\rm d}r}  ) $ \\
    $\kappa_3$ & $\frac{{M}}{{{\hbar ^2}}}{\mathop{\rm Im}\nolimits} ( \int_0^\infty  {\varphi_{s}^{(0)*}({r})\varphi_{s}^{(1)}(r)U_I({r}){\rm d}r}  )$ & $\frac{{M}}{{{\hbar ^2}}}{\mathop{\rm Im}\nolimits} ( \int_0^\infty  {\varphi_{s}^{(0)*}({r})\varphi_{s}^{(1)}(r)U_I({r})r{\rm d}r}  )$ & $\frac{{M}}{{{\hbar ^2}}}{\mathop{\rm Im}\nolimits} ( \int_0^\infty  {\varphi_{s}^{(0)*}({r})\varphi_{s}^{(1)}(r)U_I({r})r^2{\rm d}r}  )$ \\
    $\kappa_1$ & ${\rm Im}(-a_0 )$ & ${\rm Im}({\rm ln}(1/a_0) )$ & ${\rm Im}(1/a_0 )$ \\
    $\kappa_1$ & ${\rm Im}(1/a_{l>0} )$ & ${\rm Im}(1/a_{|l|>0} )$ & ${\rm Im}(1/a_{l\geq0} )$ \\
    $\kappa_2$ & ${\rm{Im}}( -{r_{l>0}^e}/2 )$ & ${\rm{Im}}( -{r_{|l|>0}^e}/2 )$ & ${\rm{Im}}( -{r_{l\geq0}^e}/2 )$ \\
    $\kappa_3$ & $\int_0^{r_0} \{[{\rm Im}\tilde\varphi_{l>0}^{(0)}(r)]^2-[{\rm Im} \varphi_{l>0}^{(0)}(r)]^2\} {\rm d}r $ & $\int_0^{r_0} \{[{\rm Im}\tilde\varphi_{|l|>0}^{(0)}(r)]^2-[{\rm Im} \varphi_{|l|>0}^{(0)}(r)]^2\}r {\rm d}r $ & $\int_0^{r_0} \{[{\rm Im}\tilde\varphi_s^{(0)}(r)]^2-[{\rm Im} \varphi_s^{(0)}(r)]^2\} r^2{\rm d}r $ \\
    \hline
  \end{tabular}
\end{table*}

\begin{table*}
  \centering
  \caption{The effective range $r_l^e$ in different dimensions \cite{Hammer2009}.}\label{table3}
  \begin{tabular}{@{}lcccc@{}}
    \hline\hline
      Dimension & $l=1$&\quad\quad& $l>1$ \\
    \hline
    1D & $r_1^e=r_0 - r_0^2 \frac{1}{a_1} + \frac{r_0^3}{3} \frac{1}{a_1^2} -  \int_0^{r_0} [\varphi^{(0)}_1(r)]^2 {\rm d} r$ &  \\
    2D & $r_1^e=\frac{1}{2} - \frac{r_0^2}{2} \frac{1}{a_1} +\frac{r_0^4}{16} \frac{1}{a_1^2} - \int_0^{r_0} [\varphi^{(0)}_1(r)]^2 r {\rm d} r$ && $r_{l>1}^e=-\frac{(2l-2)!!(2l-4)!!}{r_0^{2l-2}} - \frac{r_0^2}{2l} \frac{1}{a_l} +\frac{r_0^{2l+2}}{(2l)!!(2l+2)!!} \frac{1}{a_l^2} - \int_0^{r_0} [\varphi^{(0)}_l(r)]^2 r {\rm d} r$ \\
    3D  &\multicolumn{3}{c}{$r_{l\geq 0}^e=-\frac{(2l-1)!!(2l-3)!!}{r_0^{2l-1}} - \frac{r_0^2}{2l+1} \frac{1}{a_l} +\frac{r_0^{2l+3}}{(2l+1)!!(2l+3)!!} \frac{1}{a_l^2} - \int_0^{r_0} [\varphi^{(0)}_l(r)]^2 r^2 {\rm d} r$}\\
    \hline
  \end{tabular}
\end{table*}

\section*{III. UNIVERSAL RELATIONS FOR REACTIVE MOLECULES IN ONE DIMENSIONS}
In this section, we consider single-component reactive molecules in 1D. We need to clarify that by universal relations, we mean the relations between the two-body loss and other quantities that can be expressed in terms of contacts and microscopic parameters independent of the temperature and the total particle number. As such, the meaning of universal is different from the other context, where effects dependent only on the scattering length are called universal and the effective range and other beyond-scattering-length effects are dubbed non-universal. 

We label $s=l$, ${\bf x}_i=z_i$, ${\bf x}_{ij}=z_{ij}$, ${\bf k}=k_z$, and ${\bf X}_{ij} = {\bf R}_{ij}^z$ in the following discussions. The generalized spherical harmonics in 1D is $Y_l(\hat z_{ij})=(z_{ij}/|z_{ij}|)^l/\sqrt{2}$. The two-body wavefunction $\psi_l(z_{ij};\epsilon)=\varphi_l(|z_{ij}|;\epsilon)Y_l(\hat z_{ij})$ has the universal asymptotic form when $r_0\leq |z_{ij}| \ll k_F^{-1}$, which is
\begin{equation}
\begin{split}
\varphi_l(|z_{ij}|;\epsilon)  \stackrel{r_0\leq |z_{ij}|\ll k_F^{-1}}{\xrightarrow{\hspace*{1.6cm}} }  \frac{q_\epsilon^{l-1}}{\tan \eta_l} [&\cos(q_\epsilon |z_{ij}|-\frac{l\pi}{2})\\
&-\tan \eta_l \sin(q_\epsilon |z_{ij}|-\frac{l\pi}{2})],
\end{split}\label{asy1D}
\end{equation}
where $\eta_l$ is the 1D $l$-th partial wave phase shift and can be expanded under the low energy limit $q_\epsilon r_0 \ll 1$, as shown in Table \ref{table1}. 

\begin{center}
{\bf A. Even-wave scatterings with $l=0$}
\end{center}
We first consider the even-wave scatterings with $l=0$. For even-wave scatterings, it is sufficient to take the zero energy limit, i.e., we consider only the even-wave scattering length $a_0$ and $\varphi_0^{(0)}(|z_{ij}|)$. Based on Eq. (\ref{asy1D}), we obtain
\begin{equation}
\varphi_0^{(0)}(|z_{ij}|)  \stackrel{r_0\leq |z_{ij}|\ll k_F^{-1}}{\xrightarrow{\hspace*{1.6cm}} } a_0 - |z_{ij}| .\label{asy1D00}
\end{equation}
By taking Eq. (\ref{asy1D00}) into Eq. (\ref{asyd1}) firstly, and then bringing Eq. (\ref{asyd1}) back to Eqs. (\ref{decayd}), (\ref{MDd}) and (\ref{DCf}) respectively, We obtain the following universal relations.

\noindent {\it Two-body inelastic loss rate:} From the calculation of Eq. (\ref{decayd}), we obtain that
\begin{equation}
\frac{\partial N}{\partial t} =-\frac{\hbar}{2M} {\rm Im} (-a_0) C_1^{(0)}, \label{decay1D0}
\end{equation}
where $C_1^{(0)}$ is the 1D even-wave contact defined in Table \ref{table2}.

\noindent {\it Momentum distribution:} From Eq. (\ref{MDd}), we obtain that
\begin{equation}
n(k_z) \stackrel{k_F \leq |k_z|\ll r_0^{-1}}{\xrightarrow{\hspace*{1.6cm}} }  \frac{C_1^{(0)}}{|k_z|^4} \left|Y_0(\hat k_z)\right|^2.\label{MD1D0}
\end{equation}

\noindent {\it Density correlation function:} From Eq. (\ref{DCf}), we obtain
\begin{equation}
S(z_{ij}) \stackrel{r_0\leq |z_{ij}|\ll k_F^{-1}}{\xrightarrow{\hspace*{1.6cm}} }  \frac{1}{4} |a_0|^2 C_1^{(0)} \left|Y_0(\hat z_{ij})\right|^2.\label{DCf1D0}
\end{equation}

\begin{center}
{\bf B. Odd-wave scatterings with $l=1$}
\end{center}
Next, we consider the odd-wave scatterings with $l=1$. Based on Eq. (\ref{asy1D}), we obtain
\begin{eqnarray}
&&\varphi_1^{(0)}(|z_{ij}|)  \stackrel{r_0\leq |z_{ij}|\ll k_F^{-1}}{\xrightarrow{\hspace*{1.6cm}} } 1-\frac{|z_{ij}|}{a_1},  \label{asy1D10} \\
&&\varphi_1^{(1)}(|z_{ij}|)  \stackrel{r_0\leq |z_{ij}|\ll k_F^{-1}}{\xrightarrow{\hspace*{1.6cm}} } r_1^e |z_{ij}| -\frac{|z_{ij}|^2}{2} + \frac{1}{a_1} \frac{|z_{ij}|^3}{6},   \label{asy1D11}
\end{eqnarray}
where $a_1$ and $r_1^e$ are the 1D odd-wave scattering length and effective range, respectively. By taking Eqs. (\ref{asy1D10}) and (\ref{asy1D11}) into Eq. (\ref{asyd1}) firstly, and then bringing Eq. (\ref{asyd1}) back to Eqs. (\ref{decayd}), (\ref{MDd}) and (\ref{DCf}) respectively, We obtain the following universal relations.

\noindent {\it Two-body inelastic loss rate:} As shown in Sec. II, to calculate Eq. (\ref{decayd}), we could calculate Eqs. (\ref{decayTB}) and (\ref{decayTB1}) first. From the calculation of Eq. (\ref{decayTB1}), we have (See Appendix A)
\begin{equation}
\begin{split}
&\left.\varphi^*_1(|z_{ij}|;\epsilon) \frac{\partial}{\partial |z_{ij}|}\varphi_1(|z_{ij}|;\epsilon) \right|_{|z_{ij}|=r_0}\\
&\qquad\quad-\left. \varphi_1(|z_{ij}|;\epsilon) \frac{\partial}{\partial |z_{ij}|}\varphi^*_1(|z_{ij}|;\epsilon) \right|_{|z_{ij}|=r_0}\\
=&-\frac{1}{a_1}+\frac{1}{a^*_1} +r_1^{e}q_\epsilon^2-r_1^{e*}(q_\epsilon^2)^*+\Big[ -r_0\\
&+\frac{r_0}{2} (\frac{1}{a_1}+\frac{1}{a^*_1}) -\frac{r_0^3}{3}\frac{1}{a_1a_1^*} \Big] [q_\epsilon^2-(q_\epsilon^2)^*]+{\rm O}(q_\epsilon^4).
\end{split} \notag
\end{equation}
By using a trick that the first term on the right-hand side of the 1D odd-wave effective range in Table \ref{table3} can be rewritten as
\begin{equation}
\begin{split}
r_0=& \frac{r_1^e + r_1^{e*}}{2} + \frac{r_0^2}{2}(\frac{1}{a_1}+\frac{1}{a_1^*})-\frac{r_0^3}{6} (\frac{1}{a_1^2}+\frac{1}{(a_1^*)^2})\\
&+ \frac{1}{2}\int_0^{r_0} \{[\varphi_1^{(0)}(r)]^2+[\varphi_1^{(0)\ast}(r)]^2\} {\rm d} r ,
\end{split}
\end{equation}
one has
\begin{equation}
\begin{split}
&\left.\varphi^*_1(|z_{ij}|;\epsilon) \frac{\partial}{\partial |z_{ij}|}\varphi_1(|z_{ij}|;\epsilon) \right|_{|z_{ij}|=r_0}\\
&\qquad\quad-\left. \varphi_1(|z_{ij}|;\epsilon) \frac{\partial}{\partial |z_{ij}|}\varphi^*_1(|z_{ij}|;\epsilon) \right|_{|z_{ij}|=r_0}\\
=&-\frac{1}{a_1}+\frac{1}{a^*_1} +\frac{1}{2}(r_1^{e}-r_1^{e*})[q_\epsilon^2+(q_\epsilon^2)^*] \\
& -2\int_0^{r_0} \{[{\rm Im}\tilde\varphi_1^{(0)}(r)]^2+\frac{[\varphi_1^{(0)}(r)]^2+[\varphi_1^{(0)\ast}(r)]^2}{4}\} {\rm d} r \\
&\times[q_\epsilon^2-(q_\epsilon^2)^*]+{\rm O}(q_\epsilon^4),
\end{split} \notag
\end{equation}
where $\tilde\varphi_1^{(0)}(r)$ is obtained by extending the universal asymptotic form of $\varphi_1^{(0)}(r)$ in the range $r_0\leq r \ll k_F^{-1}$ into the range $r<r_0$, i.e., $\tilde\varphi_1^{(0)}(r) = 1-r/a_1$. Following the procedure given in Sec. II, Eq. (\ref{decayd}) can be written as
\begin{equation}
\frac{\partial N}{\partial t} =-\frac{\hbar}{2M}\sum_{\nu=1}^3 \kappa_\nu C_\nu^{(1)}, \label{decay1D1}
\end{equation}
where $C_\nu^{(1)}$ are the 1D odd-wave contacts that fully capture the many-body physics and $\kappa_\nu$ are the microscopic parameters determined purely by the two-body short range physics. Both $C_\nu^{(1)}$ and $\kappa_\nu$ are defined in Table \ref{table2}. $\kappa_1$ and $\kappa_2$ can simply be expressed as ${\rm Im}(1/a_1)$ and ${\rm Im}(-r_1^e/2)$, respectively. $\kappa_3$, however, is a new microscopic parameter emerged in the system with inelastic losses. As shown in Table \ref{table2}, the physical meaning of $\kappa_3$ is the integration of $[{\rm Im} \tilde\varphi_1^{(0)}(r)]^2$ subtracted by $[{\rm Im} \varphi_1^{(0)}(r)]^2$ with respect to $r$ from 0 to $r_0$. We note that $\kappa_2$ is the real part of the integral regarding $\varphi^{(0)*}\varphi^{(1)}U_I$, while $\kappa_3$ is the imaginary part of this integral. In general, the real and imaginary parts of $\varphi^{(0)*}\varphi^{(1)}U_I$ are independent functions. $\kappa_2$ and $\kappa_3$ are thus independent parameters.  Whereas it is possible to express $\kappa_3$ in terms of other microscopic parameters charactering the two-body interactions, here, we keep $\kappa_3$ in the expression of the universal relation, since all these parameters, $\kappa_{1,2,3}$, are independent on the particle number and temperature. As such, $\kappa_{1,2,3}$ are measurable quantities.

\noindent {\it Momentum distribution:} From Eq. (\ref{MDd}), we obtain that
\begin{equation}
n(k_z) \stackrel{k_F \leq |k_z|\ll r_0^{-1}}{\xrightarrow{\hspace*{1.6cm}} }  \frac{C_1^{(1)}}{|k_z|^2} \left|Y_1(\hat k_z)\right|^2.\label{MD1D1}
\end{equation}

\noindent {\it Density correlation function:} From Eq. (\ref{DCf}), we obtain
\begin{equation}
\begin{split}
&S(z_{ij}) \stackrel{r_0\leq |z_{ij}|\ll k_F^{-1}}{\xrightarrow{\hspace*{1.6cm}} }  \frac{1}{4} \Big[C_1^{(1)} \\
&+ \Big({\rm Re} (r_1^e) C_2^{(1)} -{\rm Re} (\frac{2}{a_1}) C_1^{(1)} -{\rm Im} (r_1^e) C_3^{(1)}\Big) |z_{ij}|  \\
&+\Big(\frac{1}{|a_1|^2} C_1^{(1)} - {\rm Re} (\frac{1}{2} + \frac{r_1^e}{a_1^*}) C_2^{(1)} + {\rm Im}(\frac{r_1^e}{a_1^*})C_3^{(1)}\Big)|z_{ij}|^2\\
&+\Big({\rm Re} (\frac{2}{a_1})C_2^{(1)} + {\rm Im} (\frac{1}{a_1}) C_3^{(1)} \Big) \frac{|z_{ij}|^3}{3}\\
&-\frac{1}{|a_1|^2} C_2^{(1)}\frac{|z_{ij}|^4}{6} \Big]\left|Y_1(\hat z_{ij})\right|^2.
\end{split}\label{DCf1D1}
\end{equation}

By fitting the data of quantities such as the momentum distribution and the density correlation function obtained in experiment, all quantities in Eqs. (\ref{decay1D0}) and (\ref{decay1D1}) can be measured.

For s-wave inelastic scatterings, only the leading term determined by scattering length $a_0$ is important in the low energy expansion of phase shift (See Table \ref{table1}). $C_1^{(1)}$ alone is enough to describe physics in such systems. Generally speaking, for high-partial wave scatterings with a generic short-range interaction, other microscopic parameters like the effective range are required in the low energy expansion of phase shift. The low energy expansion of the wavefunction needs to be kept up to the $q_\epsilon^2$ term in Eq. (\ref{asyd1}). As such, all three contacts $C_\nu^{(1)}$ are required in the complete expressions of the universal relations that apply to all parameter regimes. Nevertheless, in certain parameter regimes, the terms including $C_2^{(1)}$ and $C_2^{(1)}$ in Eq. (\ref{decay1D1}) may be less important. For instance, in weakly interacting systems when $a_1\to 0$, the contribution to the phase shift is dominated by the scattering length and other microscopic parameters can be neglected. As such, the universal relations are mainly governed by $C_1^{(1)}$, similar to the original universal relations for s-wave scatterings. This could simplify data analysis in experiments as fewer parameters are required to fit the experimental results. This argument works for 2D and 3D systems as well.

\section*{IV. UNIVERSAL RELATIONS FOR REACTIVE MOLECULES IN TWO DIMENSIONS}
We now consider single-component reactive molecules in 2D.  We label $s=l$, ${\bf x}_i={\bm \rho}_i$, ${\bf x}_{ij}={\bm \rho}_{ij}=(x_{ij},y_{ij})$ and ${\bf X}_{ij} = {\bf R}_{ij}^{\bm \rho}$ in the following discussions. The generalized spherical harmonics in 2D is $Y_l(\hat {\bm \rho}_{ij})=[(x_{ij}+iy_{ij})/\rho_{ij}]^l/\sqrt{2\pi}$, where $\rho_{ij}=|{\bm \rho}_{ij}|$. The two-body wavefunction $\psi_l({\bm \rho}_{ij};\epsilon)=\varphi_l(\rho_{ij};\epsilon)Y_l(\hat {\bm \rho}_{ij})$ has the universal asymptotic form when $r_0\leq \rho_{ij} \ll k_F^{-1}$, which is
\begin{equation}
\begin{split}
\varphi_l(\rho_{ij};\epsilon)  \stackrel{r_0\leq \rho_{ij}\ll k_F^{-1}}{\xrightarrow{\hspace*{1.6cm}} }  \frac{\pi}{2}\frac{q_\epsilon^l}{\tan \eta_l} [&J_l(q_\epsilon \rho_{ij}) \\
&-\tan \eta_l N_l (q_\epsilon \rho_{ij})],
\end{split}\label{asy2D}
\end{equation}
where $J_l$ ($N_l$) is the Bessel function of the first (second) kind and $\eta_l$ is the 2D $l$-th partial wave phase shift and can be expanded under the low energy limit $q_\epsilon r_0 \ll 1$, as shown in Table \ref{table1}.

\begin{center}
{\bf A. s-wave scatterings with $l=0$}
\end{center}
We first consider the s-wave scatterings with $l=0$.  For s-wave scatterings, it is sufficient to take the zero energy limit, i.e., we consider only the s-wave scattering length $a_0$ and $\varphi_0^{(0)}(\rho_{ij})$. We obtain
\begin{equation}
\varphi_0^{(0)}(\rho_{ij})  \stackrel{r_0\leq \rho_{ij}\ll k_F^{-1}}{\xrightarrow{\hspace*{1.6cm}} } \ln a_0 - \ln \rho_{ij} .\label{asy2D00}
\end{equation}
By taking Eq. (\ref{asy2D00}) into Eq. (\ref{asyd1}) firstly, and then bringing Eq. (\ref{asyd1}) back to Eqs. (\ref{decayd}), (\ref{MDd}) and (\ref{DCf}) respectively, We obtain the following universal relations.

\noindent {\it Two-body inelastic loss rate:} From the calculation of Eq. (\ref{decayd}), we obtain that
\begin{equation}
\frac{\partial N}{\partial t} =-\frac{\hbar}{2\pi^2M} {\rm Im} (\ln \frac{1}{a_0}) C_1^{(0)}, \label{decay2D0}
\end{equation}
where $C_1^{(0)}$ is the 2D s-wave contact defined in Table \ref{table2}.

\noindent {\it Momentum distribution:} From Eq. (\ref{MDd}), we obtain that
\begin{equation}
n({\bf k}_{\bm\rho}) \stackrel{k_F \leq |{\bf k}_{\bm\rho}|\ll r_0^{-1}}{\xrightarrow{\hspace*{1.6cm}} }  \frac{C_1^{(0)}}{|{\bf k}_{\bm\rho}|^4} \left|Y_0(\hat {\bf k}_{\bm\rho})\right|^2.\label{MD2D0}
\end{equation}

\noindent {\it Density correlation function:} From Eq. (\ref{DCf}), we obtain
\begin{equation}
S({\bm \rho}_{ij}) \stackrel{r_0\leq \rho_{ij}\ll k_F^{-1}}{\xrightarrow{\hspace*{1.6cm}} }  \frac{1}{(2\pi)^2} |\ln \rho_{ij}|^2 C_1^{(0)} \left|Y_0(\hat {\bm \rho}_{ij})\right|^2.\label{DCf2D0}
\end{equation}

\begin{center}
{\bf B. High partial wave scatterings with $l>0$}
\end{center}
Next, we consider the high partial wave scatterings with $l>0$. From Eq. (\ref{asy2D}), we obtain
\begin{eqnarray}
&&\varphi_{l>0}^{(0)}(\rho_{ij})  \stackrel{r_0\leq \rho_{ij}\ll k_F^{-1}}{\xrightarrow{\hspace*{1.6cm}} } -\frac{1}{a_l} \frac{\rho_{ij}^l}{(2l)!!}+\frac{(2l-2)!!}{\rho_{ij}^l},  \label{asy2D10} \\
&&\varphi_{l=1}^{(1)}(\rho_{ij})  \stackrel{r_0\leq \rho_{ij}\ll k_F^{-1}}{\xrightarrow{\hspace*{1.6cm}} } r_1^e \frac{\rho_{ij}}{2} -\ln (\frac{\rho_{ij}}{r_0})\frac{\rho_{ij}}{2} + \frac{1}{a_1} \frac{\rho_{ij}^3}{16}, \notag \\
&& \\  \label{asy2D11}
&&\varphi_{l>1}^{(1)}(\rho_{ij})  \stackrel{r_0\leq \rho_{ij}\ll k_F^{-1}}{\xrightarrow{\hspace*{1.6cm}} } r_l^e \frac{\rho_{ij}^l}{(2l)!!} +\frac{1}{a_l}\frac{\rho_{ij}^{l+2}}{2(2l+2)!!} \notag \\
&&\qquad\qquad\qquad\qquad\qquad  + \frac{(2l-4)!!}{2} \frac{1}{\rho_{ij}^{l-2}},   \label{asy2D12}
\end{eqnarray}
where $a_l$ and $r_l^e$ are the 2D $l$-th partial wave scattering length and effective range, respectively. By taking Eqs. (\ref{asy2D10}-\ref{asy2D12}) into Eq. (\ref{asyd1}) firstly, and then bringing Eq. (\ref{asyd1}) back to Eqs. (\ref{decayd}), (\ref{MDd}) and (\ref{DCf}) respectively, We obtain the following universal relations.

\noindent {\it Two-body inelastic loss rate:} As shown in Sec. II, to calculate Eq. (\ref{decayd}), we could calculate Eqs. (\ref{decayTB}) and (\ref{decayTB1}) first. From the calculation of Eq. (\ref{decayTB1}), we have (See Appendix A)
\begin{equation}
\begin{split}
&\left.\rho_{ij}\varphi^*_1(\rho_{ij};\epsilon) \frac{\partial}{\partial \rho_{ij}}\varphi_1(\rho_{ij};\epsilon) \right|_{\rho_{ij}=r_0}\\
&\qquad\quad-\left. \rho_{ij}\varphi_1(\rho_{ij};\epsilon) \frac{\partial}{\partial \rho_{ij}}\varphi^*_1(\rho_{ij};\epsilon) \right|_{\rho_{ij}=r_0}\\
=&-\frac{1}{a_1}+\frac{1}{a^*_1} +r_1^{e}q_\epsilon^2-r_1^{e*}(q_\epsilon^2)^*+\Big[ -\frac{1}{2}\\
&+\frac{r_0^2}{4} (\frac{1}{a_1}+\frac{1}{a^*_1}) -\frac{r_0^4}{16}\frac{1}{a_1a_1^*} \Big] [q_\epsilon^2-(q_\epsilon^2)^*]+{\rm O}(q_\epsilon^4)
\end{split} \notag
\end{equation}
for the p-wave scattering and
\begin{equation}
\begin{split}
&\left.\rho_{ij}\varphi^*_l(\rho_{ij};\epsilon) \frac{\partial}{\partial \rho_{ij}}\varphi_l(\rho_{ij};\epsilon) \right|_{\rho_{ij}=r_0}\\
&\qquad\quad-\left. \rho_{ij}\varphi_l(\rho_{ij};\epsilon) \frac{\partial}{\partial \rho_{ij}}\varphi^*_l(\rho_{ij};\epsilon) \right|_{\rho_{ij}=r_0}\\
=&-\frac{1}{a_l}+\frac{1}{a^*_l} +r_l^{e}q_\epsilon^2-r_l^{e*}(q_\epsilon^2)^*+\Big[ \frac{(2l-2)!!(2l-4)!!}{r_0^{2l-2}}\\
&+\frac{r_0^2}{4l} (\frac{1}{a_l}+\frac{1}{a^*_l}) -\frac{r_0^{2l+2}}{(2l)!!(2l+2)!!}\frac{1}{a_la_l^*} \Big] [q_\epsilon^2-(q_\epsilon^2)^*]\\
&+{\rm O}(q_\epsilon^4)
\end{split} \notag
\end{equation}
for the higher partial wave scatterings with $l>1$. By using the same trick in 1D case that the first term on the right-hand side of the 2D effective range in Table \ref{table3} can be rewritten as
\begin{equation}
\begin{split}
\frac{1}{2}=& \frac{r_1^e + r_1^{e*}}{2} + \frac{r_0^2}{4}(\frac{1}{a_1}+\frac{1}{a_1^*})-\frac{r_0^4}{32} (\frac{1}{a_1^2}+\frac{1}{(a_1^*)^2})\\
&+ \frac{1}{2}\int_0^{r_0} \{[\varphi_1^{(0)}(r)]^2+[\varphi_1^{(0)\ast}(r)]^2\} r {\rm d} r
\end{split}
\end{equation}
for the p-wave scattering and
\begin{equation}
\begin{split}
&-\frac{(2l-2)!!(2l-4)!!}{r_0^{2l-2}}= \frac{r_l^e + r_l^{e*}}{2} + \frac{r_0^2}{4l}(\frac{1}{a_l}+\frac{1}{a_l^*}) \\
&\qquad\qquad\quad -\frac{r_0^{2l+2}}{2(2l)!!(2l+2)!!} (\frac{1}{a_l^2}+\frac{1}{(a_l^*)^2})\\
&\qquad\qquad\quad + \frac{1}{2}\int_0^{r_0} \{[\varphi_l^{(0)}(r)]^2+[\varphi_l^{(0)\ast}(r)]^2\} r {\rm d} r 
\end{split}
\end{equation}
for the higher partial wave scatterings with $l>1$, one has
\begin{equation}
\begin{split}
&\left.\rho_{ij}\varphi^*_l(\rho_{ij};\epsilon) \frac{\partial}{\partial \rho_{ij}}\varphi_l(\rho_{ij};\epsilon) \right|_{\rho_{ij}=r_0}\\
&\qquad\quad-\left. \rho_{ij}\varphi_l(\rho_{ij};\epsilon) \frac{\partial}{\partial \rho_{ij}}\varphi^*_l(\rho_{ij};\epsilon) \right|_{\rho_{ij}=r_0}\\
=&-\frac{1}{a_l}+\frac{1}{a^*_l} +\frac{1}{2}(r_l^{e}-r_l^{e*})[q_\epsilon^2+(q_\epsilon^2)^*] \\
& -2\int_0^{r_0} \{[{\rm Im} \tilde\varphi_l^{(0)}(r)]^2 +\frac{[\varphi_l^{(0)}(r)]^2+[\varphi_l^{(0)\ast}(r)]^2}{4}\} r {\rm d} r \\
&\times[q_\epsilon^2-(q_\epsilon^2)^*]+{\rm O}(q_\epsilon^4)
\end{split} \notag
\end{equation}
for $l>0$, where $\tilde\varphi_l^{(0)}(r)$ is obtained from extending the universal asymptotic form of $\varphi_l^{(0)}(r)$ in the range $r_0\leq r \ll k_F^{-1}$ into the range $r<r_0$, i.e., $\tilde\varphi_{l>0}^{(0)}(\rho_{ij}) =-(1/a_l)[ \rho_{ij}^l/(2l)!!]+(2l-2)!!/\rho_{ij}^l$. It is interest to notice that, for high partial wave scatterings, the above formula suits for 3D systems as well, which gives
\begin{equation}
\begin{split}
&\left.|{\bf x}_{ij}|^{d-1}\varphi^*_l(|{\bf x}_{ij}|;\epsilon) \frac{\partial}{\partial |{\bf x}_{ij}|}\varphi_l(|{\bf x}_{ij}|;\epsilon) \right|_{|{\bf x}_{ij}|=r_0}\\
&\qquad\quad-\left. |{\bf x}_{ij}|^{d-1}\varphi_l(|{\bf x}_{ij}|;\epsilon) \frac{\partial}{\partial |{\bf x}_{ij}|}\varphi^*_l(|{\bf x}_{ij}|;\epsilon) \right|_{|{\bf x}_{ij}|=r_0}\\
=&-\frac{1}{a_l}+\frac{1}{a^*_l} +\frac{1}{2}(r_l^{e}-r_l^{e*})[q_\epsilon^2+(q_\epsilon^2)^*] \\
& -2\int_0^{r_0} \{[{\rm Im} \tilde\varphi_l^{(0)}(r)]^2 +\frac{[\varphi_l^{(0)}(r)]^2+[\varphi_l^{(0)\ast}(r)]^2}{4}\} r^{d-1} {\rm d} r \\
&\times[q_\epsilon^2-(q_\epsilon^2)^*]+{\rm O}(q_\epsilon^4).
\end{split} \notag
\end{equation}

Following the procedure given in Sec. II, Eq. (\ref{decayd}) can be written as
\begin{equation}
\frac{\partial N}{\partial t} =-\frac{\hbar}{2\pi^2M}\sum_{\nu=1}^3 \kappa_\nu C_\nu^{(l)}, \label{decay2D1}
\end{equation}
where $C_\nu^{(l)}$ are the 2D $l$-th partial wave contacts that fully capture the many-body physics and $\kappa_\nu$ are the microscopic parameters determined purely by the two-body short range physics. Both $C_\nu^{(l)}$ and $\kappa_\nu$ are defined in Table \ref{table2}. $\kappa_1$ and $\kappa_2$ can simply be expressed as ${\rm Im}(1/a_l)$ and ${\rm Im}(-r_l^e/2)$, respectively. Similar to that for 1D odd-wave scatterings, $\kappa_3$, however, is a new microscopic parameter emerged in the system with inelastic losses. As shown in Table \ref{table2}, again, the physical meaning of $\kappa_3$ is the integration of $[{\rm Im} \tilde\varphi_l^{(0)}(r)]^2$ subtracted by $[{\rm Im} \varphi_l^{(0)}(r)]^2$ with respect to $r$ from 0 to $r_0$.

\noindent {\it Momentum distribution:} From Eq. (\ref{MDd}), we obtain that
\begin{equation}
n({\bf k}_{\bm \rho}) \stackrel{k_F \leq |{\bf k}_{\bm \rho}|\ll r_0^{-1}}{\xrightarrow{\hspace*{1.6cm}} }  C_1^{(l)}|{\bf k}_{\bm \rho}|^{2l-4} \left|Y_l(\hat {\bf k}_{\bm \rho})\right|^2.\label{MD2D1}
\end{equation}

\noindent {\it Density correlation function:} From Eq. (\ref{DCf}), we obtain
\begin{equation}
\begin{split}
&S({\bm\rho}_{ij}) \stackrel{r_0\leq \rho_{ij}\ll k_F^{-1}}{\xrightarrow{\hspace*{1.6cm}} }  \frac{1}{(2\pi)^2} \Big[C_1^{(1)}\frac{1}{\rho_{ij}^{2}} -C_2^{(1)} \frac{\ln (\rho_{ij}/r_0)}{2}\\
&+ \Big({\rm Re} (r_1^e) C_2^{(1)} -{\rm Re} (\frac{2}{a_1}) C_1^{(1)} -{\rm Im} (r_1^e) C_3^{(1)}\Big) \frac{1}{2} \\
&+\Big({\rm Re} (\frac{1}{a_1})C_2^{(1)} + {\rm Im} (\frac{1}{a_1}) C_3^{(1)} \Big) \frac{\ln (\rho_{ij}/r_0)\rho_{ij}^2}{4}\\
&+\Big(\frac{1}{|a_1|^2} C_1^{(1)} - {\rm Re} (\frac{r_1^e-1/4}{a_1^*}) C_2^{(1)} \\
&\qquad+{\rm Im}(\frac{r_1^e+1/4}{a_1^*})C_3^{(1)}\Big)\frac{\rho_{ij}^{2}}{4}\\
&-\frac{1}{|a_1|^2} C_2^{(1)}\frac{\rho_{ij}^{4}}{32} \Big]\left|Y_1(\hat {\bm \rho}_{ij})\right|^2
\end{split}\label{DCf2D1}
\end{equation}
for the p-wave scattering and
\begin{equation}
\begin{split}
&S({\bm\rho}_{ij}) \stackrel{r_0\leq \rho_{ij}\ll k_F^{-1}}{\xrightarrow{\hspace*{1.6cm}} }  \frac{1}{(2\pi)^2} \Big[C_1^{(l)}\frac{[(2l-2)!!]^2}{\rho_{ij}^{2l}} \\
&+C_2^{(l)}\frac{(2l-4)!!(2l-2)!!}{2\rho_{ij}^{2l-2}}\\
&+ \Big({\rm Re} (r_l^e) C_2^{(l)} -{\rm Re} (\frac{2}{a_l}) C_1^{(l)} -{\rm Im} (r_l^e) C_3^{(l)}\Big) \frac{1}{2l} \\
&-\Big(\frac{1}{l}{\rm Re} (\frac{1}{a_l})C_2^{(l)} + {\rm Im} (\frac{1}{a_l}) C_3^{(l)} \Big) \frac{\rho_{ij}^2}{(2l-2)(2l+2)}\\
&+\Big(\frac{1}{|a_l|^2} C_1^{(l)} - {\rm Re} (\frac{r_l^e}{a_l^*}) C_2^{(l)} + {\rm Im}(\frac{r_l^e}{a_l^*})C_3^{(l)}\Big)\frac{\rho_{ij}^{2l}}{[(2l)!!]^2}\\
&-\frac{1}{|a_l|^2} C_2^{(l)}\frac{\rho_{ij}^{2l+2}}{2(2l)!!(2l+2)!!} \Big]\left|Y_l(\hat {\bm \rho}_{ij})\right|^2
\end{split}\label{DCf2D2}
\end{equation}
for the higher partial wave scatterings with $l>1$. 

By fitting the data of quantities such as the momentum distribution and the density correlation function obtained in experiment, all quantities in Eqs. (\ref{decay2D0}) and (\ref{decay2D1}) can be measured. It is worth pointing out that it might be difficult to distinguish certain terms such as ${\ln (\rho_{ij}/r_0)\rho_{ij}^2}$ and $\ln (\rho_{ij}/r_0)$ in practice. It is nevertheless useful to keep the full expression of the universal relation as a complete description, which shall be useful even for purely theoretical studies. In experiments, despite that the full expression may lead to difficulties in fitting the experimental data, a unique feature is that the same universal relation applies to both the weakly and strongly interacting regimes and also any particle numbers. Furthermore, in certain parameter regimes, some terms may be more important than others. For instance, when $C_2^{(1)}\ll {\rm Re} (1/a_1)C_2^{(1)} + {\rm Im} (1/a_1) C_3^{(1)}$, the term dependent on ${\ln (\rho_{ij}/r_0)\rho_{ij}^2}$  shall be more important than that dependent on $\ln (\rho_{ij}/r_0)$. This may simplify the fitting procedures.

\section*{V. DISCUSSION}
In Table \ref{table2}, we list the two-body inelastic loss rate in 1D, 2D, and 3D. One can recognize that the two-body inelastic loss rate has exactly the same form in all $d$D, which is
\begin{equation}
\frac{\partial N}{\partial t} = -\frac{2\hbar}{\Omega_d^2M} \sum_{\nu=1}^3 \kappa_\nu C_\nu^{(s)},\label{decaydg}
\end{equation}
or, equivalently, 
\begin{equation}
\frac{\partial n}{\partial t} = -\frac{2\hbar}{\Omega_d^2M} \sum_{\nu=1}^3 \kappa_\nu {\cal C}_\nu^{(s)}\label{decaydgd}
\end{equation}
where $\Omega_d$ is the solid angle in $d$D, which is $\Omega_1=2$, $\Omega_2=2\pi$, and $\Omega_3=4\pi$, respectively. $n=N/L_d$ and ${\cal C}_\nu^{(s)}=C_\nu^{(s)}/L_d$ are the molecular density and the contact density of the system, respectively. $L_d$ is the size of the system in $d$D. Whereas $\kappa_\nu$ behave very differently in different dimensions for s-wave (even-wave for 1D) scatterings, which is originated from the distinct behavior of the low energy expansion of the phase shift in different dimensions as shown in Table \ref{table1}, they are exactly the same for high partial wave scatterings, regardless of the dimension of the system.

We need to emphasize that we have considered the short-range interactions $U({\bf r})$ with a cut-off length $r_0$ throughout this work to demonstrate the physics underlying the universal relations in lossy systems at low dimensions. When an electric field is applied, the dipole moment of a polar molecule becomes finite,  and the dipole-dipole interaction $\sim A/|{\bf r}|^n$ with $n=3$ would become important. Generally, for dilute systems with the power-law interaction $\sim A/|{\bf r}|^n$ where $n>2$, a characteristic length $\tilde r=(M|A|/\hbar^2)^{1/(n-2)}$ can be defined \cite{Friedrich2013}. When $\tilde r \ll |{\bf r}|\ll k_F^{-1}$, due to such a length scale separation, the many-body wavefunction has universal asymptotic behavior Eq. (\ref{asyd1}) as well \cite{Hofmann2021}. Following the method presented in our manuscript, universal relation Eq. (\ref{decaydg}) can also be obtained. While the low energy expansion of the phase shift might be very different that the scattering length and effective range may not be well defined \cite{OMalley1961,Gao2008}, new microscopic parameters determined by the details of the interactions, such as $n$ and $l$, needs to be used. For instance, without losses, universal relations for systems with dipole-dipole interactions has been studied \cite{Hofmann2021}. In lossy systems like reactive molecules, it will be interesting to study how the power-law interactions influence contacts, universal relations and the decay rates.

Equations (\ref{decaydg}) and (\ref{decaydgd}) are exact for any many-body eigenstates. Thus, it is invariant under the thermal average. 

\begin{center}
{\bf A. Temperature dependence of the loss rate in homogeneous systems}
\end{center}
We take a two-body system in free space as an example. In this case, $\epsilon$ becomes a good quantum number. The two-body wavefunction can be written as $\Psi({\bf x}_1,{\bf x}_2)=\phi_{\rm c} ({\bf X}_{12})\psi_s({\bf x}_{12})$, where $\phi_{\rm c} ({\bf X}_{12})$ is the normalized wavefunction of the center of mass motion of the two molecules. $\psi_s({\bf x}_{12})$ is 
\begin{equation}
\begin{split}
\psi_l(z_{12})=\Big[\sqrt{\frac{2\Omega_1}{L_1}} &\frac{q_\epsilon^l}{q_\epsilon^{2l-1}(\cot \eta_l -i)} \Big] \frac{q_\epsilon^{l-1}}{\tan \eta_l}[ \cos(q_\epsilon |z_{12}|-\frac{l\pi}{2}) \\
&-\tan \eta_l\sin(q_\epsilon |z_{12}|-\frac{l\pi}{2})]Y_l(\hat z_{12})
\end{split}\label{TB1D}
\end{equation}
for 1D systems and
\begin{equation}
\begin{split}
\psi_l({\bm \rho}_{12})= \Big[\sqrt{\frac{2\Omega_2}{ L_2}}  &\frac{q_\epsilon^{l}}{(\pi/2)q_\epsilon^{2l}(\cot \eta_l -i)} \Big] \frac{\pi}{2} \frac{q_\epsilon^l}{\tan \eta_l} [J_l(q_\epsilon \rho_{12})\\
&-\tan \eta_l N_l(q_\epsilon \rho_{12})]Y_l(\hat {\bm \rho}_{12})
\end{split}\label{TB2D}
\end{equation}
for 2D systems, where $L_1$ is the length of the 1D system and $L_2$ is the area of the 2D system. Recall that, in 3D, 
\begin{equation}
\begin{split}
\psi_{lm}({\bf r}_{12})= \Big[ \sqrt{\frac{2\Omega_3}{L_3}}  &\frac{q_\epsilon^l}{q_\epsilon^{2l+1}(\cot \eta_l -i)} \Big] \frac{q_\epsilon^{l+1}}{\tan \eta_l} [j_l(q_\epsilon |{\bf r}_{12}|)\\
&-\tan \eta_l n_l(q_\epsilon |{\bf r}_{12}|)]Y_{lm}(\hat {\bf r}_{12}),
\end{split}\label{TB3D}
\end{equation}
where ${\bf r}_{12}=({\bm \rho}_{12},z_{12})$, $j_l$ ($n_l$) is the spherical Bessel function of the first (second) kind, and $L_3$ is the volume of the 3D system.

By denoting $C_\nu^{[l]}$ as $C_\nu^{(l)}$ for 1D, $ C_\nu^{(l)}+C_\nu^{(-l)}$ for 2D, and $\sum\nolimits_{m}C_\nu^{(lm)}$ for 3D, respectively, and based on the definition in Table \ref{table2}, $C_\nu^{[l]}$ in $d$D is expressed as
\begin{eqnarray}
C_1^{[l]}&=& \sigma_d\frac{4\Omega_d^3}{L_d}  \left|q_\epsilon^l f_{l,d} (q_\epsilon) \right|^2, \label{C1d} \\
C_2^{[l]}&=&2{\rm Re}(q_\epsilon^2)C_1^{[l]}, \label{C2d} \\
C_3^{[l]}&=&2{\rm Im}(q_\epsilon^2)C_1^{[l]}, \label{C3d}
\end{eqnarray}
where $f_{l,d} (q_\epsilon)\equiv1/\{[1+\delta_{d,2}(\pi/2-1)]q_\epsilon^{2l+d-2}(\cot \eta_l -i)\}$ and $\delta_{d,d'}$ is the Kronecker delta. $\sigma_d$ is the fold of degeneracy for the $l$-th partial wave scatterings in $d$D, which is $\sigma_1=1$, $\sigma_2=2$, and $\sigma_3=2l+1$, respectively. Based on the results shown in Table \ref{table1}, $f_{l,d} (q_\epsilon)$ can be expanded in the low-energy limit, $f_{l,d} (q_\epsilon)=f_{l,d}^{(0)}+{\rm O} (q_\epsilon)$, where $f_{l,d}^{(0)}$ is $q_\epsilon$ independent and relates only to the scattering length $a_l$. Note that $f_{0,2}^{(0)}=1/[\ln (a_0e^\gamma/2)]$. As an example, we consider the scattering states only and the case that only the term $f_{l,d}^{(0)}$ in $f_{l,d} (q_\epsilon)$ is important, where $q_\epsilon$ can treated as a real quantity and $C_3^{[l]}=0$. 

By considering the second-order virial expansion only and based on the two-body results as shown in Eqs. (\ref{C1d}-\ref{C3d}), the thermal averaged contacts can be obtained by doing the calculation \cite{He2020}
\begin{equation}
\langle C_\nu ^{[l]} \rangle_T=Z^{-1} e^{\frac{2\mu}{k_B T}} \sum_{E_c} e^{-\frac{E_c}{k_B T}}\sum_{n} C_\nu ^{[l]}e^{-\frac{\epsilon_n}{k_B T}},\label{CdT0}
\end{equation}
where $Z$ is the partition function, $E_c=\hbar^2q_c^2/(4M)$ is the energy of the center of mass motion with momentum $q_c$, $\epsilon_n=\hbar^2q_{\epsilon_n}^2/M$ is the eigenenergy of the relative motion with momentum $q_{\epsilon_n}$, and $k_B$ is the Boltzmann constant. $\mu$ is the chemical potential, which can be extracted from $N=k_B T \partial_\mu \ln Z$. In the high temperature regime, $N/L_d\approx\exp[\mu/(k_B T)]/\lambda_T^d$, where $\lambda_T=[2\pi\hbar^2/(k_BTM)]^{1/2}$ is the thermal wavelength. We have
\begin{equation}
\langle C_\nu ^{[l]} \rangle_T=2^{\frac{d}{2}-1} N^2 \lambda_T^d \frac{\Omega_d}{(2\pi)^d} \int_0^\infty C_\nu^{[l]} \exp(-\frac{\lambda_T^2}{2\pi}k^2) k^{d-1}{\rm d}k. \label{CdT}
\end{equation}
Thus, we obtain $\langle C_\nu ^{[l]} \rangle_T$ as a function of $N$ and $T$ by substituting Eqs. (\ref{C1d}) and (\ref{C2d}) into Eq. (\ref{CdT}). Based on the fact that $\int_0^\infty \exp(-nx^2) x^{{\rm w}-1} {\rm d} x= 2^{-1}\Gamma({\rm w}/2)n^{-{\rm w}/2}$, we obtain
\begin{eqnarray}
\langle C_1 ^{[l]} \rangle_T & = & 2^l \pi^{l-\frac{d}{2}} \Gamma(l+\frac{d}{2}) \sigma_d \Omega_d^4  \left|f_{l,d}^{(0)}\right|^2 \frac{N^2}{L_d} \lambda_T^{-2l}, \label{C1dt}\\
\langle C_2 ^{[l]} \rangle_T & = & 2^{l+2} \pi^{l-\frac{d}{2}+1} \Gamma(l+\frac{d}{2}+1) \sigma_d \Omega_d^4 \notag\\
&&\qquad\qquad\quad\quad\times\left|f_{l,d}^{(0)}\right|^2 \frac{N^2}{L_d} \lambda_T^{-2l-2}. \label{C2dt}
\end{eqnarray}

\begin{center}
{\bf B. Temperature dependence of the loss rate in harmonic traps}
\end{center}
When a harmonic trap, $V_{\rm ext}({\bf x})=(1/2) M\omega^2({\bf x}\cdot{\bf x})$, is applied, under the local density approximation, we can replace $\mu$ by the local chemical potential $\mu({\bf x})=\mu (0)- V_{\rm ext}({\bf x})$ and write $n({\bf x})=\exp[\mu({\bf x})/(k_B T)]/\lambda_T^d$ in the high temperature regime. $\omega$ is the harmonic frequency. $\mu (0)$ is the chemical potential at the center of the trap. At any point ${\bf x}$ in the trap, Eq. (\ref{decaydgd}) still applies, we have
\begin{equation}
\frac{\partial n({\bf x})}{\partial t} = -\frac{2\hbar}{\Omega_d^2M} \sum_{\nu=1}^3 \kappa_\nu \langle{\cal C}_\nu^{[l]} ({\bf x})\rangle_T, \label{decaydgdt}
\end{equation}
Thus, by taking the integration over ${\bf x}$ on both sides of Eq. (\ref{decaydgdt}), the two-body inelastic loss rate in $d$D traps can be written as
\begin{equation}
\frac{\partial N^{\rm trap}}{\partial t} = -\frac{2\hbar}{\Omega_d^2M} \sum_{\nu=1}^3 \kappa_\nu \langle C_\nu^{[l]\rm trap} \rangle_T. \label{decaydgt}
\end{equation}
Based on Eqs. (\ref{C1dt}) and (\ref{C2dt}), at the center of the trap, we still have
\begin{eqnarray}
\langle {\cal C}_1 ^{[l]} (0) \rangle_T & = & 2^l \pi^{l-\frac{d}{2}} \Gamma(l+\frac{d}{2}) \sigma_d \Omega_d^4  \notag\\
&&\qquad\qquad\quad\quad\times\left|f_{l,d}^{(0)}\right|^2 n^2(0) \lambda_T^{-2l}, \label{C1dt0}\\
\langle {\cal C}_2 ^{[l]} (0) \rangle_T & = & 2^{l+2} \pi^{l-\frac{d}{2}+1} \Gamma(l+\frac{d}{2}+1) \sigma_d \Omega_d^4 \notag\\
&&\qquad\qquad\quad\quad\times\left|f_{l,d}^{(0)}\right|^2 n^2(0) \lambda_T^{-2l-2},\label{C2dt0}
\end{eqnarray}
where $n(0)$ can be expressed by $N^{\rm trap}$ and $T$, which is 
\begin{equation}
N^{\rm trap}=\int n({\bf x}) {\rm d}{\bf x}=\left(\frac{2\pi k_B T}{M\omega^2}\right)^{d/2}n(0).\label{Ntrapc}
\end{equation}
The total contacts, $\langle C_\nu^{\rm trap}\rangle_T$, can be determined by integrating the local contacts in the trap,
\begin{equation}
\begin{split}
\langle C_\nu^{[l]\rm trap}\rangle_T=&\langle {\cal C}_{\nu} ^{[l]} (0) \rangle_T \int e^{-2V_{\rm ext} ({\bf x})/(k_B T)}{\rm d} {\bf x} \\
=&\left(\frac{\pi k_B T}{M\omega^2}\right)^{d/2}\langle {\cal C}_{\nu} ^{[l]} (0) \rangle_T.
\end{split}\label{CdTt}
\end{equation}

Thus, based on Eqs. (\ref{decaydgt}-\ref{CdTt}), one can map the loss rate in a harmonic trap to the one in a homogeneous system by setting the effective size of the homogeneous system to be $\tilde L_d=[4\pi k_B T/(M\omega^2)]^{d/2}$. One has
\begin{equation}
\frac{\partial N^{\rm trap}}{\partial t} = - \beta_{l,d} \frac{(N^{\rm trap})^2}{\tilde L_d}\label{decaydgte0}
\end{equation}
or, equivalently, 
\begin{equation}
\frac{\partial \tilde n}{\partial t} = - \beta_{l,d} \tilde n^2\label{decaydgte1}
\end{equation}
where $\tilde n=N^{\rm trap}/\tilde L_d$ is the average molecular density of the system and $\beta_{l,d}$ is the loss rate coefficient for $d$D $l$-th partial wave scatterings.

\section*{VI. CONCLUSION}
In conclusion, we have established universal relations for the two-body inelastic loss rate, which are controlled by contacts $C_\nu^{(s)}$ in 1D and 2D, respectively. Whereas $\kappa_\nu$ have different forms in different dimensions for s-wave (even-wave for 1D) scatterings, the loss rate can be written as exactly the same form in arbitrary dimensions for high partial wave (odd-wave for 1D) scatterings. Moreover, the two-body inelastic loss rate can be related to other physical quantities such as the momentum distribution and the density correlation function through contacts. While we considered single-component ultracold atoms or reactive molecules, discussions can be generalized to multi-component systems straightforwardly. It will also be interesting to consider a finite confinement in the transverse direction such that the dimension crossover can be explored in the presence of two-body losses.  We hope that our work could inspire more efforts of using contacts and universal relations to study novel phenomena in lossy quantum systems in condensed matter physics, atomic, molecular and optical physics, and chemical physics.

\section*{ACKNOWLEDGEMENTS}
M.H. is supported by the Shenzhen Science and Technology Program (JCYJ20210324140805014) and acknowledges the financial support from NSAF U1930402 and computational resources from the Beijing Computational Science Research Center. Q.Z. acknowledges support from NSF through PHY 2110614.

\renewcommand{\theequation}{A\arabic{equation}}
\setcounter{equation}{0}
\section*{APPENDIX A:  MATHEMATICS USED IN THE CALCULATION OF EQUATION (\ref{decayTB1}) }

For a specific partial wave scattering only, to calculate Eq. (\ref{decayTB1}), one can first calculate
\begin{equation}
\begin{split}
&\int_0^{r_0} {\rm d} {\bf x}_{ij}\Big[ \psi^*_s (  {\bf x}_{ij};\epsilon)\nabla^2_{{\bf x}_{ij}} \psi_s (  {\bf x}_{ij};\epsilon) \\
&\qquad\qquad-\psi_s (  {\bf x}_{ij};\epsilon)\nabla^2_{{\bf x}_{ij}} \psi^*_s (  {\bf x}_{ij};\epsilon)  \Big]\\
=&\int_0^{r_0} {\rm d} {\bf x}_{ij}\Big[ \varphi^*_s(|{\bf x}_{ij}|;\epsilon) Y^*_{\{0\}} (\hat {\bf x}_{ij})\nabla^2_{{\bf x}_{ij}} \varphi_s(|{\bf x}_{ij}|;\epsilon) Y_{\{0\}} (\hat {\bf x}_{ij}) \\
&\qquad-\varphi_s(|{\bf x}_{ij}|;\epsilon) Y_{\{0\}} (\hat {\bf x}_{ij})\nabla^2_{{\bf x}_{ij}} \varphi^*_s(|{\bf x}_{ij}|;\epsilon) Y^*_{\{0\}} (\hat {\bf x}_{ij})  \Big]\\
=&\frac{1}{\Omega_d}\oint_{|{\bf x}_{ij}|=r_0} \Big[ \varphi^*_s(|{\bf x}_{ij}|;\epsilon) \frac{\partial}{\partial|{\bf x}_{ij}|} \varphi_s(|{\bf x}_{ij}|;\epsilon)  \\
&\qquad\qquad\qquad-\varphi_s(|{\bf x}_{ij}|;\epsilon) \frac{\partial}{\partial|{\bf x}_{ij}|} \varphi^*_s(|{\bf x}_{ij}|;\epsilon)   \Big] \hat {\bf e}_{\bf x}\cdot{\rm d} {\bf S}.
\end{split} \label{A1}
\end{equation}
where $Y_{\{0\}}=Y_{s=\{0\}}$ means that all the quantum numbers in $s$ are zero. $\hat {\bf e}_{\bf x}$ is the outgoing unit vector perpendicular to ${\bf S}$. $|Y_{\{0\}}|^2=1/\Omega_d$ is also used. Thus, to calculate Eq. (\ref{decayTB1}), it is helpful to first calculate
\begin{equation}
\begin{split}
&\left.|{\bf x}_{ij}|^{d-1}\varphi^*_s(|{\bf x}_{ij}|;\epsilon) \frac{\partial}{\partial |{\bf x}_{ij}|}\varphi_s(|{\bf x}_{ij}|;\epsilon) \right|_{|{\bf x}_{ij}|=r_0}\\
&\qquad\quad-\left. |{\bf x}_{ij}|^{d-1}\varphi_s(|{\bf x}_{ij}|;\epsilon) \frac{\partial}{\partial |{\bf x}_{ij}|}\varphi^*_s(|{\bf x}_{ij}|;\epsilon) \right|_{|{\bf x}_{ij}|=r_0}.
\end{split} \label{A2}
\end{equation}

\end{document}